# Laser-Pumped Drilling Carbon Nanotube Vortex Shock Waves in Optical Fibers

**Ricardo E. da Silva,[1] and Marcos A. R. Franco[2]**

[1]Institute of Physics Gleb Wataghin, University of Campinas (UNICAMP), Campinas, 13083-859, Brazil
[2]Institute for Advanced Studies (IEAv), São José dos Campos, 12245-021, Brazil

Corresponding author: Ricardo E. da Silva (e-mail: resilva@unicamp.br).

This work was supported in part by the grant 2022/10584-9, São Paulo Research Foundation (FAPESP). We thank the LIMicro-IQ – Microscopy Core Facility (RRID:SCR_024633) at UNICAMP for the support (the Quanta FEG 250 system was partially funded by a FAPESP grant (#2023/01620-4)), and the Laboratório de Plasmas e Processos (LPP) from Instituto Tecnológico de Aeronáutica, Brazil, for providing the analysis on the Profilometer KLA Tencor P-7 (software Profiler 8.0). The Article Processing Charges for the publication of this research was funded by the Coordenação de Aperfeiçoamento de Pessoal de Nível Superior - CAPES identifier ROR: 00x0ma614.

**ABSTRACT** We experimentally demonstrate laser-induced vortex shock waves formed by carbon nanotubes drilling optical fibers for the first time. Three samples of standard single-mode optical fibers (SMF) are individually inserted in a syringe loaded with a 1 mL solution of single-walled carbon nanotubes (CNT) and methanol, and a high-power laser is injected into the fibers for 5 (SMF 1), 10 (SMF 2), and 20 (SMF 3) minutes. The CNT solution thermally expands and generates vortex acoustic flows, which are confined in the syringe cavity, significantly increasing the velocity and impact of nanotubes at the fiber tip. The resulting shock waves achieve estimated hypersonic velocities (5742 m/s) and high pressures (6.7 GPa), overcoming the silica tensile strength and ablating structured vortices in the fibers. The material, geometry, and depth profile of the vortices are characterized, providing details of mixing carbon and silica layers, increasing radially from the fiber core center and in thickness to the cladding for longer laser periods (850 nm to 10 μm thickness). The cross-sections of the measured vortices are compared to analytical simulations, revealing unprecedented Fibonacci helices drilling holes in the fiber core with a 5 μm maximum depth, while depositing nanoscale CNT-silica layers following Fibonacci spirals. These results reveal novel contributions compared to previous studies, such as fabrication of nano-to-micro scale CNT-silica Fibonacci vortices in SMFs with controlled radial and thickness dimensions, laser-induced CNT shock waves with significantly higher velocities, pressures, and ablating properties, and a highly efficient and low-cost syringe technique. These achievements are significant to deposit other nanoparticles besides CNTs (e.g., silver and gold) and fabrication of all-fiber vortex spatial phase modulators for optical communications, quantum optics, micromanipulation, fiber sensors for chemical, biological, and gas substances, efficient modulation of mode-locked fiber lasers using evanescent fields, and high-resolution biomedical ultrasonic neurotransmitters.

## I. INTRODUCTION

Carbon nanotubes (CNT) integrated in optical fibers provide outstanding optical, chemical, and mechanical features for fiber sensors, optical switches, fiber lasers, and biomedical ultrasonic devices [1], [2], [3], [4], [5]. CNTs offer a wide wavelength absorption range (400 - 1200 nm) with peaks tuned by the tube diameter and chirality [6], [7], covering the broad spectrum of active doped fibers (e.g., Ytterbium, Erbium, and Thulium) to modulate fiber lasers with femtosecond pulse durations and kW power outputs [7], [8], [9], [10]. Thus, the thickness, shape, and composition of CNT layers deposited on the fiber tips impact the output laser properties and damage threshold [10]. For example, thick CNT-polymer layers deposited in the fiber core (> 50 μm thickness) improve the laser modulation depth while reducing the phase noise. In contrast, the resulting increased non-saturable losses lower the laser power output and damage threshold [10]. Alternatively, depositing CNTs around the fiber core (ring-shaped layer) favors the overlapping of optical evanescent fields and nanoparticles, improving device lifetime and overall laser power output [1]. For fiber-based ultrasonic applications, CNT-polymer composites deposited at the fiber tip can generate high-pressure ultrasonic waves to treat dementia, Alzheimer's, and depression [5]. A pulsed laser is injected into the fiber, exciting the CNT composite, which thermally expands, focusing ultrasonic waves on brain tissues to stimulate a single neuron or group of neurons [5]. CNT layers with increasing thickness are usually employed to generate high-pressure ultrasound, while thin layers are suitable for increasing frequency and bandwidth, improving spatial resolution in biomedical imaging sensors and neurostimulators [5], [11].

The CNT deposition on optical fibers has been successfully demonstrated employing wallpaper techniques [10], chemical functionalization with organic gels and motorized dip coating [12], and using laser radiation [2], [4], [9], [13], [14], resulting in layer thickness decreasing from ~ 49 μm to 900 nm [10], [12], [15]. Laser-induced deposition provides a fast and controllable spatial distribution of CNT bundles over the fiber cross-section [2], [4], [13], favoring the use of simpler CNT solutions based on water and alcohols [4]. In those experiments, the optical fiber is vertically dipped in a CNT solution while the laser is injected, attracting nanoparticles toward the fiber tip. The deposited layers on the fiber core, around the core, and the overall fiber cross-section can be controlled by changing the fiber type, laser power level, wavelength, deposition time, CNT, and solvent properties [1], [4]. Nevertheless, the particles' motion to the fiber is usually restricted inside the laser beam cone, depending on a competitive combination of complex physical mechanisms: fluid thermal expansion (thermophoresis) and convection moving CNTs to the laser beam axis in converging trajectories; and strong radiation pressure diverging and expelling CNTs away from the fiber core [6], [14]. Although thermophoretic flows have been demonstrated, fiber-driven vortex shock waves composed of CNTs have not yet been reported. Numerical simulations have shown that carbon nanotubes absorbing laser energy can thermally expand and heat the surrounding fluid (which also expands), generating overlapped acoustic waves [16]. Particle thermal expansion is the primary mechanism for laser-driven generation of shock waves using optical fibers employing dye absorbers [17], [18], [19].

Here, we demonstrate the laser-driven generation, development, and impact of vortex shock waves in a CNT-methanol solution, ablating carbon-silica vortices on the tip of optical fibers. The theoretical background, shock wave properties, and analytical simulation of Fibonacci vortices are reviewed in Section II. Three SMF samples are investigated for increasing laser durations (5, 10, and 20 minutes, respectively) using a new syringe-based cavity method, as described in Section III. Section IV-A shows a detailed characterization of the material, geometric, and thickness properties of the CNT-silica structures fabricated on the SMFs' cross-section. The shock waves are characterized, and the shock velocities and pressures are estimated and discussed. The measured CNT-silica vortices are compared to the simulations (Section IV-B), revealing remarkable Fibonacci spirals inscribed by the shock waves, which are promising for a new broad demand of vortex-based integrated photonic and acoustic devices (Section V).

## II. THEORETICAL BACKGROUND

### A. SHOCK WAVE GENERATION

Laser-induced shock waves generated by optical fibers have been demonstrated using fluid solutions composed of dye optical absorbers dispersed in water [17], [18], [19]. In general, the dye absorbs the laser radiation, converting a portion of the energy into acoustic waves, which superpose and expand from the fiber tip. Thus, the shock wave magnitude attenuates with decreasing power of the diverging laser cone from the fiber. Consequently, the decreasing acoustic expansion of shock waves can usually be observed in a maximum region of about 600 μm around the fiber tip [17], [18], [19].

Fig. 1 illustrates the laser-induced generation of CNT vortex shock waves from the tip of a standard single-mode optical fiber (SMF) employing similar physical mechanisms [17], [18], [19]. Fig. 1(a) shows the cross-section of a single-walled nanotube (CNT) generating acoustic waves in a fluid as a mechanical piston, absorbing laser power and thermally expanding (dashed black), and heating the fluid that also thermally expands (dashed orange) [16]. CNT and fluid waves mostly overlap in phase (blue curves), contributing to the overall acoustic flow. Fig. 1(b) shows the formation of a shock wave when CNTs surrounding the fiber are exposed to the laser under power absorption saturation [18]. The heated particles around the fiber tip combine as an acoustic source diffracting thermoelastic waves: compressive pressure waves (orange plane wave) conforming to the fiber shape; and toroidal tensile waves (blue circles) radiated by the fiber edges [17], [18], [19]. The toroidal waves overlap, creating a vortex with a maximum tensile pressure $P$ along the fiber axis, rupturing the fluid and forming a cloud of cavitation bubbles [19]. Fig. 1(c) illustrates a nanoscale cross-section of a bubble around an ideal bundle of nanotubes, which usually form long lengths compared to its diameter (randomly distributed bundles in the fluid might agglomerate, forming clusters with distinct shapes and sizes) [14], [20], [21], [22]. The bubbles expand and collapse, contributing to accelerating CNTs to the fiber tip [23]. Fig. 1(d)-(f) illustrate the expansion of a vortex shock wave for increasing arbitrary laser durations ($t_1$, $t_2$, $t_3$). The accumulating CNT bubbles along the fiber axis increase the shock wave dimensions [19]. Consequently, CNTs that hit and attach to the fiber tip will form deposited layers with increasing diameter and thickness.

Here, we show that vortex shock waves composed of CNTs are strongly confined and amplified inside a syringe-based cylindrical cavity. Fig. 1(g) illustrates the container cross-section (with a length $H$ and radius $R$) of a syringe loaded with a CNT-fluid solution, indicating the formation of shock waves under laser radiation from the SMF. The fiber radiates acoustic flows that expand and propagate in the solution inside and outside the laser beam [18]. After a period of laser exposure, the toroidal waves reach and conform to the container walls, being guided and confined inside the cylinder [24], [25]. The toroidal waves overlap and generate a vortex that moves CNT bundles in helical trajectories, accumulating along the fiber axis mostly in the laser cone due to higher power absorption. Fig. 1(h) illustrates an ideal cross-section of an arbitrary vortex shock wave with tip diameter decreasing towards the fiber core as a funnel [17], [18], [19]. The vortex tip can be composed of multiple azimuthally spaced helices merging in a unique helical flow with a maximum pressure $P$ at the core axis [26], [27], [28], [29] (e.g., as illustrated in the vortex core with two helices in the inset in Fig. 1(h)).

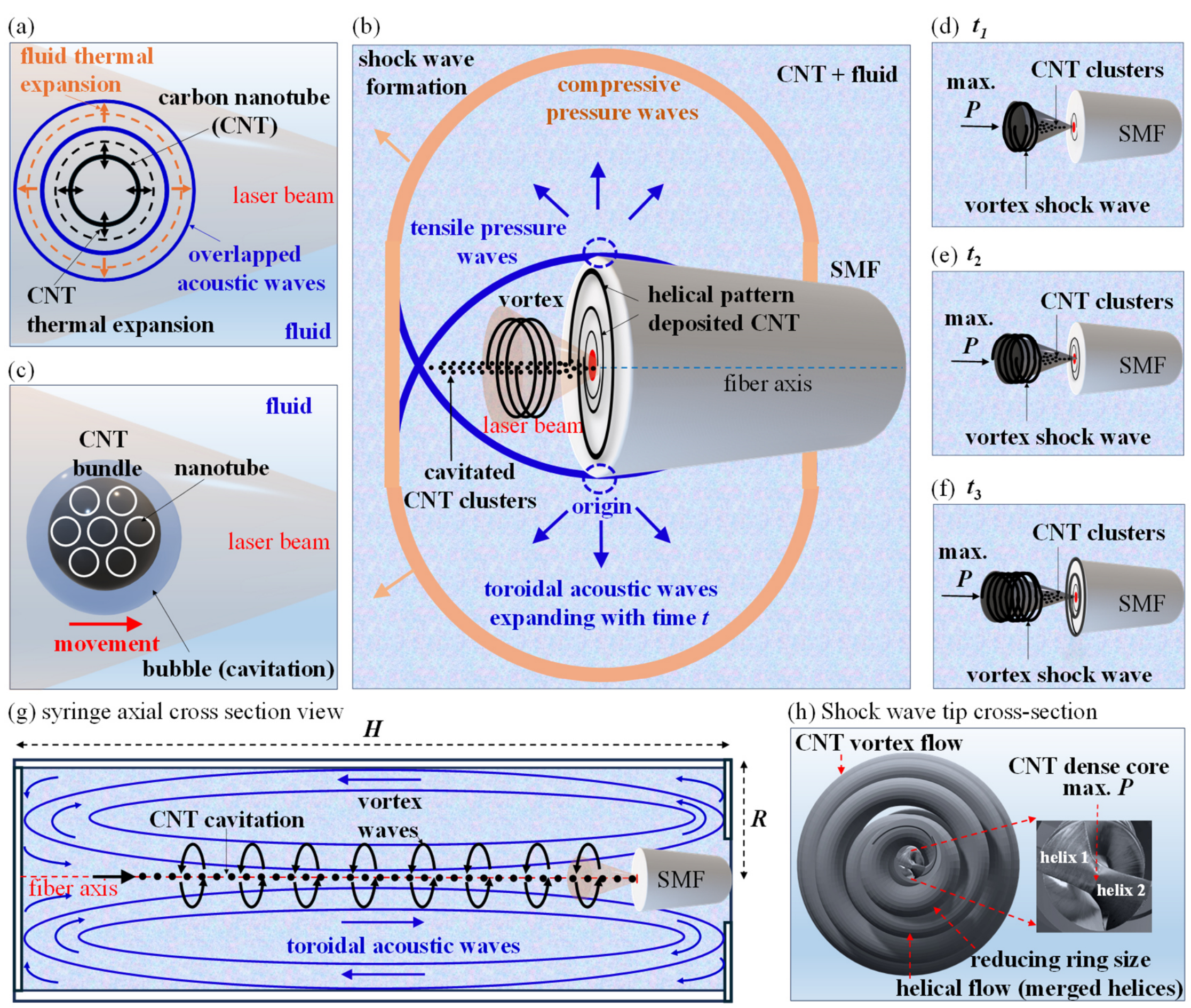

**FIGURE 1.** Illustration of a standard single-mode optical fiber (SMF) injecting a laser beam in a fluid solution with carbon nanotubes (CNT) generating vortex shock waves. (a) Cross-section of a nanotube thermally expanding and heating the surrounding fluid, which also thermally expands, generating acoustic waves. (b) SMF tip works as an acoustic source diffracting compressive pressure waves (orange) and toroidal tensile waves (blue), which overlap, inducing a vortex with maximum tensile pressure $P$ along the fiber axis. The vortex tensile waves rupture the fluid, causing (c) cavitation bubbles around CNT bundles. (d)-(f) CNT cavitation increases for prolonged arbitrary laser duration times, $t_1$, $t_2$, $t_3$. (g) Confinement and reinforcement of vortex shock waves inside a cylindrical container with length $H$ and radius $R$ inside a syringe. (h) Cross-section front of an arbitrary vortex shock wave, indicating a dense core of nanotubes.

### B. ABLATING SHOCK WAVE PROPERTIES AND VORTEX ANALYTICAL MODELING

Fig. 2(a) illustrates a vortex along the $z$ axis generating a CNT helical flow converging with the radius $r$ to the core center ($r = 0$) after a full angular rotation ($\theta$) ($\Lambda$ is the spatial period [30]). For an ideal compound vortex, the vortex rotates with angular velocity $\Omega$ and distinct azimuthal velocities, $V_\theta = \Omega r$ ($r < R_C$), and $V_\theta = \Omega R_C^2/r$ ($r > R_C$), where $R_C$ is the vortex core radius (approximated to the fiber core radius $R_C$). For short laser duration, CNTs concentrate mostly on the surrounding helices, moving with increasing axial velocities and being gradually deposited over the SMF cross-section (the CNT bundles are indicated with solid black circles in Fig. 2(a)). The deposited CNT bundles are ideally unaffected by the following colliding particles from the vortex at this stage [31], [32]). For long laser duration, CNTs cavitating along the fiber axis can overlap as a large and elongated bubble, as illustrated in Fig. 2(b). This bubble collects CNTs in the fluid and accelerates the particles as a "jet" to the fiber core [23], [33], [34] (the vortex helices are omitted, and the yellow regions indicate power emitted by the nanotubes in the visible spectrum).

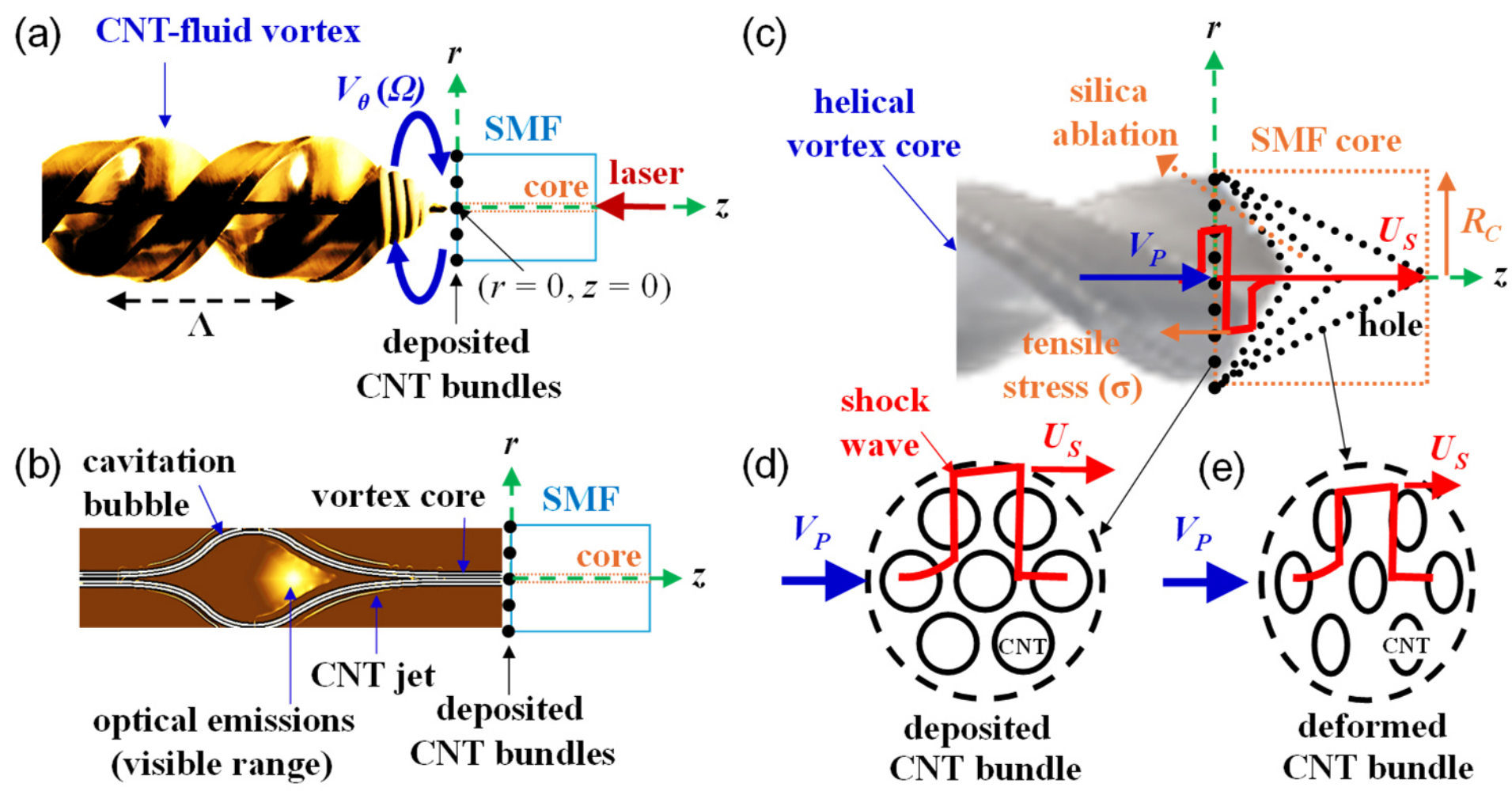

**FIGURE 2.** (a) Illustration of an arbitrary vortex along the $z$ axis radially expanding from the fiber core center ($r = 0$) and rotating with the azimuthal $V_\theta$, and angular velocity $\Omega$. (b) cavitation bubble jetting carbon nanotubes (CNT) to the tip of a standard single-mode fiber (SMF). (c) Detail of a dense vortex core with particles traveling with velocity $V_P$ colliding with CNT bundles deposited on the fiber core of radius $R_C$ for a (d) short and (e) long laser periods. The resulting shock waves traveling with velocity $U_S$ are partially reflected at the silica interface, overcoming its tensile stress $\sigma$ and ablating the material while depositing new CNT layers in the fiber core.

For a cylindrical cavity with length-to-radius ratio, $H/R \geq 4$ [25], the rotation of the vortex core for the bubble near the fiber tip can be described in terms of the Reynolds number, estimated as,

$$Re_B = 817H - 217, \tag{1}$$

and [25], [34],

$$Re_B = \frac{\rho_P \Omega R_C^2}{\mu_P}, \tag{2}$$

where [35], [36],

$$\rho_P = (1-\eta)\rho_F + \eta\rho_{CNT}, \tag{3}$$

is the effective density of the CNT-fluid solution, depending respectively on the fluid and CNT densities, $\rho_F$, and $\rho_{CNT}$, and the CNT volume concentration in the solution, $\eta$. The effective solution viscosity is [21], [37],

$$\mu_P = (1 + 13.5\eta + 904.4\eta^2)\mu_F, \tag{4}$$

where, $\mu_F$, is the dynamic fluid viscosity. The axial velocity of CNTs colliding at the fiber core axis ($r = 0$, $z = 0$) is estimated as [26],

$$V_P = \sqrt{2}V_{\theta max} \tag{5}$$

where, $V_{\theta\,max} = \Omega R_C$. The ratio of the azimuthal $V_{\theta\,max}$ and axial $V_P$ velocities defines the theoretical threshold of the swirl parameter, $S_c \sim 2V_{\theta max}/V_P = \sqrt{2} = 1.41$ [26].

Fig. 2(c) illustrates the vortex core ablating silica from the SMF core for increasing duration of laser exposure. The CNT-fluid particles colliding with the axial velocity $V_P$ at the CNT bundles deposited on the silica interface (with nano thickness and negligible losses) induce shock waves, which propagate in the nanotube arrays with the shock velocity given as [38], [39], [40],

$$U_S = \tau V_P + c_{CNT}, \tag{6}$$

where, $\tau$ is the CNT bundle Gruneisen parameter, $c_{CNT} = \sqrt{E_{CNT}/\rho_{CNT}}$, is the sound speed in an ideally undeformed nanotube bundle (Fig. 2(d)), $E_{CNT}$ and $\rho_{CNT}$, are respectively the bundle effective Young's modulus and density [41]. Thus, the pressure induced by the shock waves in the CNT bundles is [38], [39], [40],

$$P_S = \rho_{CNT} V_P U_S. \tag{7}$$

For particle velocities in the range of $V_P$ = 500 - 1000 m/s, inducing shock pressures up to $P_S$ = 5 – 6.6 GPa in a linear elastic deformation stage, the nanotubes experience reversal elastic compression, as illustrated in Fig. 2(e) (the nanotubes further recover, preserving the initial structure in Fig. 2(d)) [42], [43]. For increasing shock pressures, the nanotubes in the bundles might undergo buckling, developing structural transitions at extreme pressures ($P_S \sim$ 40 - 55 GPa), causing collapse, structural reconfiguration, changes in inter-tube connections, and material densification [41], [43]. Overall, increasing pressures favor the mechanical reinforcement of the resulting deposited CNT layers [35], [41].

TABLE I
PROPERTIES OF SINGLE-WALLED CARBON NANOTUBES (CNT) AND FLUID TO ESTIMATE THE PARTICLE ($V_P$) AND SHOCK WAVE ($U_S$) VELOCITIES

| Symbol | Parameter | Value | Ref. |
|---|---|---|---|
| $R$ | container radius | 5 mm | - |
| $H$ | container length | 20.5 mm | - |
| $H/R$ | length-to-radius ratio | ~ 4 | - |
| $Re_B$ | Reynolds number | 3051 | Eq. (1) |
| $\rho_F$ | methanol density | 784.59 kg/m$^3$ | [44] |
| $\rho_{CNT}$ | CNT density (bundle) | 1610 kg/m$^3$ | [41], [42] |
| $\rho_P$ | CNT-fluid density | 785.82 kg/m$^3$ | Eq. (3) |
| $\mu_F$ | methanol viscosity | 0.5294 mPa.s | [44] |
| $\mu_P$ | CNT-fluid viscosity | 0.5412 mPa.s | Eq. (4) |
| $\eta$ | CNT volume concentration (average 0.1% - 0.2%) | 0.15% | [45] |
| $E_{CNT}$ | CNT Young's modulus (bundle) | 37 GPa | [46] |
| $\tau$ | CNT Gruneisen parameter (average) | 1.3 | [47], [48] |
| $R_C$ | standard single-mode fiber (SMF) radius | 4.1 µm | - |

Parameters estimated from available data at $T = 300$ K.

The CNT shock waves are partially reflected on the silica interface, inducing a negative wave phase due to distinct CNT-silica acoustic impedances [40]. The negative wave amplitude causes a tensile stress exceeding the silica tensile strength $\sigma$, corroding and removing silica that is spread over the cladding. The ablating physical mechanisms might cause spallation, opening, and growing cracks into larger cavities and fractures. The induced tensile stress ejects silica and nanotubes due to strong volume expansion during crack growth [49], [50]. After the first layer is removed from the surface, a new free surface with CNTs is formed, gradually ablating the hole depth for longer laser duration, as illustrated in Fig. 1(c).

Previous studies have shown that vortices develop three azimuthal helices converging to the vortex core axis for the range of $S_C \sim 1.3$ - 1.4, and $Re = 1690$ - 3049 [26], [29] (these ranges are not strictly defined, and lower values of $S_C$ might induce one or two helices alone or superposed [27], [28]). Thus, we have experimentally set up an acoustic cylindrical cavity with radius $R$ = 5 mm and length $H$ = 20.5 mm, satisfying the condition in (1) with $H/R \sim 4$. The estimated Reynolds number $Re_B = 3051$ and the critical swirl parameter $S_C \sim 1.4$ prioritize a vortex with three helices.

The vortex cross-section is further modeled by using 2D analytical formulations developed for planar hydrodynamics, defining a generic flow induced by Fibonacci vortices [51], [52], [53]. The flow is described in the $xy$ plane by concentric circles with the radius $r$ ratio described by the Golden ratio as, $\varphi = (1 + \sqrt{5})/2 \sim 1.6$ [51]. A Fibonacci vortex is therefore defined by the rotation of the parametric Fibonacci spiral curve, as [52],

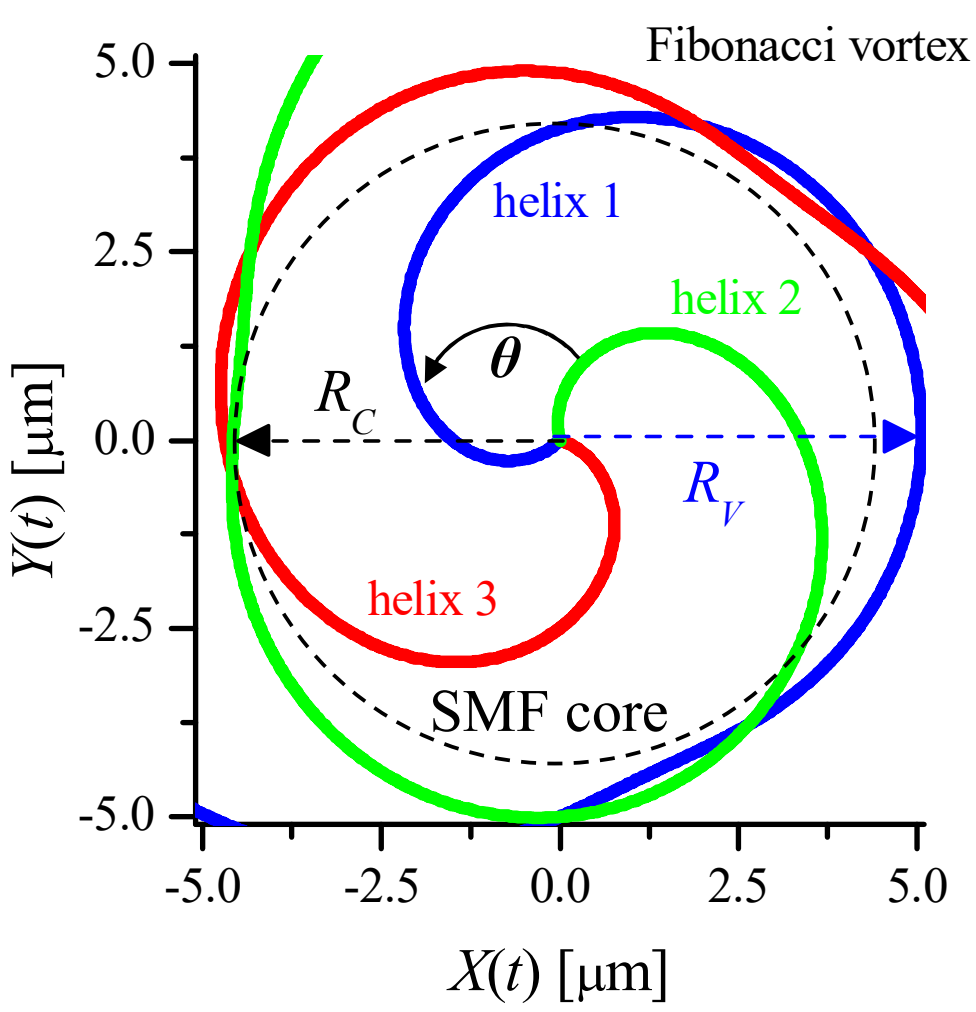

**FIGURE 3. 2D analytical simulation of a Fibonacci vortex composed by three helices converging to the vortex core axis ($R_V$ = 0). The Fibonacci spirals overlap as a ring with a maximum radius $R_V$ inscribing the standard single-mode fiber (SMF) core with radius $R_C$.**

$$x(t) = \frac{1}{\sqrt{5}}\left[e^{t \ln \varphi} \cos\big(\phi(t-1)\big) - e^{-t \ln \varphi} \cos\big(\pi t + \phi(t-1)\big)\right], \tag{8}$$

$$y(t) = \frac{1}{\sqrt{5}}\left[e^{t \ln \varphi} \sin\big(\phi(t-1)\big) - e^{-t \ln \varphi} \sin\big(\pi t + \phi(t-1)\big)\right], \tag{9}$$

where $t$ is the range corresponding to the number of sectors in the $2\pi$ azimuthal plane, increasing with the Fibonacci sequence, and $\Phi$ is the effective topological charge in each section, determining the rotation of the spiral in the $t$ range [54]. The parameters in (8) and (9) describe a vortex developing from a unique dimensionless Fibonacci spiral.

Here, we show that the helices in a Fibonacci shock wave vortex can be described by the superposition of three Fibonacci spirals azimuthally spaced with $\theta = 120°$ ($\Phi = 2\pi/3$) [55]. The overlapped spirals inscribe a circle with unitary radius $r$ = 1, which is scaled to fiber dimensions as $X(t) = x(t)R_V$, and $Y(t) = y(t)R_V$, where $R_V$ is the vortex radius. Fig. 3 shows the simulation of a ring formed by the three helices with a diameter $D_V = 2R_V = 10$ µm covering the SMF core ($D_C = 2R_C = 8.2$ µm) over the $xy$ plane at $z$ = 0. $R_V$ changes as a funnel in the vortex tip from the fiber axis ($R_V = 0$), and projects the 2D helices at any $z$ depth in the ablated fiber core hole. This analytical method is further used to characterize the boundaries and edges of the vortex structures inscribed by the shock waves.

The shock wave properties and velocities are estimated based on parameters available for bundles of single-walled CNTs with diameters approaching the employed 0.84 nm nanotubes, as well, for the methanol at the temperature of $T = 300$ $K$. The parameters are summarized with references and equations used in the calculations in Table 1.

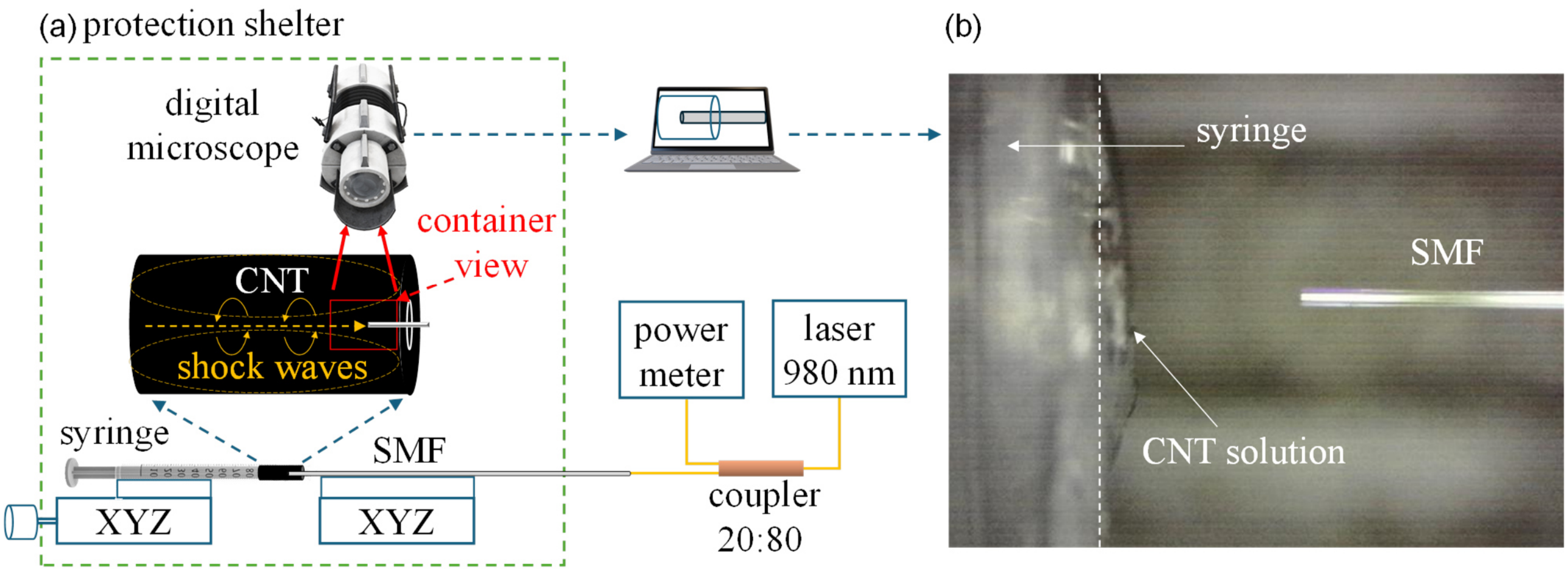

**FIGURE 4. (a) Illustration of the experimental setup employed to generate and characterize the laser-induced vortex shock waves in a solution with carbon nanotubes (CNT) dispersed in methanol. (b) The standard single-mode optical fiber (SMF) is axially aligned to the syringe's aperture and inserted in the solution by using XYZ micrometer positioners.**

## III. EXPERIMENTAL SETUP AND METHODS

Fig. 4(a) illustrates the experimental setup to generate and characterize the vortex shock waves [56]. A dispersion of single-walled carbon nanotubes (SCNT model Sigma-Aldrich 775533 with 0.84 nm average diameter and carbon purity ≥ 95 %) was added to a 15 mL polypropylene tube along with 15 mL of concentrated ammonium hydroxide solution (27 - 30 %) and kept at rest, preventing ammonia evaporation. The material was centrifuged at 3500 rpm for 5 minutes and washed. The wet solid nanotubes were added to a beaker with methanol (purity ≥ 99.8 %), being sonicated in a probe at room temperature. The mixture was filtered through paper and stored in glass vials of 20 mL. The resulting solution has a CNT concentration of ~ 2.5 mg/mL. Further details about the preparation of the CNT-methanol solution are found in [57], [58]. A syringe composed of polypropylene with a maximum capacity of 3 mL is loaded with a 1 mL CNT-methanol solution. The syringe is fixed on an XYZ micrometer positioner, as indicated in Fig. 4(a) (the detail shows the loaded container forming a cylindrical cavity of 20.5 mm long and 10 mm in diameter, with an aperture of 3.5 mm in diameter).

The fiber is cleaved, and the tip flatness is checked and cleaned by applying low-power arc voltage with a fusion splicer. SMF is also fixed on an XYZ positioner and centrally aligned to the syringe's aperture with the assistance of a digital microscope connected to a laptop. Fig. 4(b) shows the syringe aperture with the CNT solution and the SMF before inserting the tip in the solution. A removable shelter is used to block laser emissions through the syringe, additionally contributing to isolating the components and CNT solution from contamination and illumination from the environment.

The laser centered at 980 nm is split by the coupler, delivering a maximum power of about 161 mW at the SMF tip and a low-power signal to a power meter monitoring the power output. The generation and characterization of shock waves are investigated for three SMF samples by changing the laser duration (laser period ON): SMF 1 (5 minutes), SMF 2 (10 minutes), and SMF 3 (20 minutes), considering the same setup parameters and CNT solution. The microscope is focused on the fiber tip, and the visible radiation emitted by the shock waves is monitored in a microscope's maximum window view of ~ 4.9 x 3.6 mm, as illustrated in Fig. 4(a) (red square). The shock waves are recorded at arbitrary times for SMF 3, showing the strongest optical emissions among the samples, and indicating wave fields expanding in the fluid. The images are further analyzed with the ImageJ (FIJI) software [59].

An electron scanning microscope (SEM - FEI Quanta FEG 250) is used to image high-resolution details of the deposited CNT structures over the SMFs' cross-section and estimate the layers' lateral thickness. The layer thickness profile in the fiber core region is also measured for SMF 2 and 3 by using a profilometer (KLA Tencor P-7). Although it is not presented, we characterized the material properties of the CNT-silica structures, in different positions at the fiber's cross section, by using an energy dispersive spectrometer (EDS - Oxford X-Max 50) connected to SEM and a Raman spectrometer (micro–Raman XploRA Horiba). EDS is used to identify the material chemical elements by detecting X-rays emitted from localized regions in the structure, while the Raman spectrum indicates the specific resonances of the employed nanotubes mixed with silica. The measured compound materials, composed mostly of carbon (CNT) or carbon mixed with silica (CNT + silica), are further indicated in the figures in the next section.

## IV. RESULTS AND DISCUSSION

### A. MATERIAL AND DIMENSION PROPERTIES OF THE SHOCK WAVE INSCRIBED VORTICES

Fig. 5(a)-(c) show the CNT-silica vortices over the cross-sections of SMF 1, 2, and 3 inscribed by the shock waves during laser exposure for 5, 10, and 20 minutes, respectively.

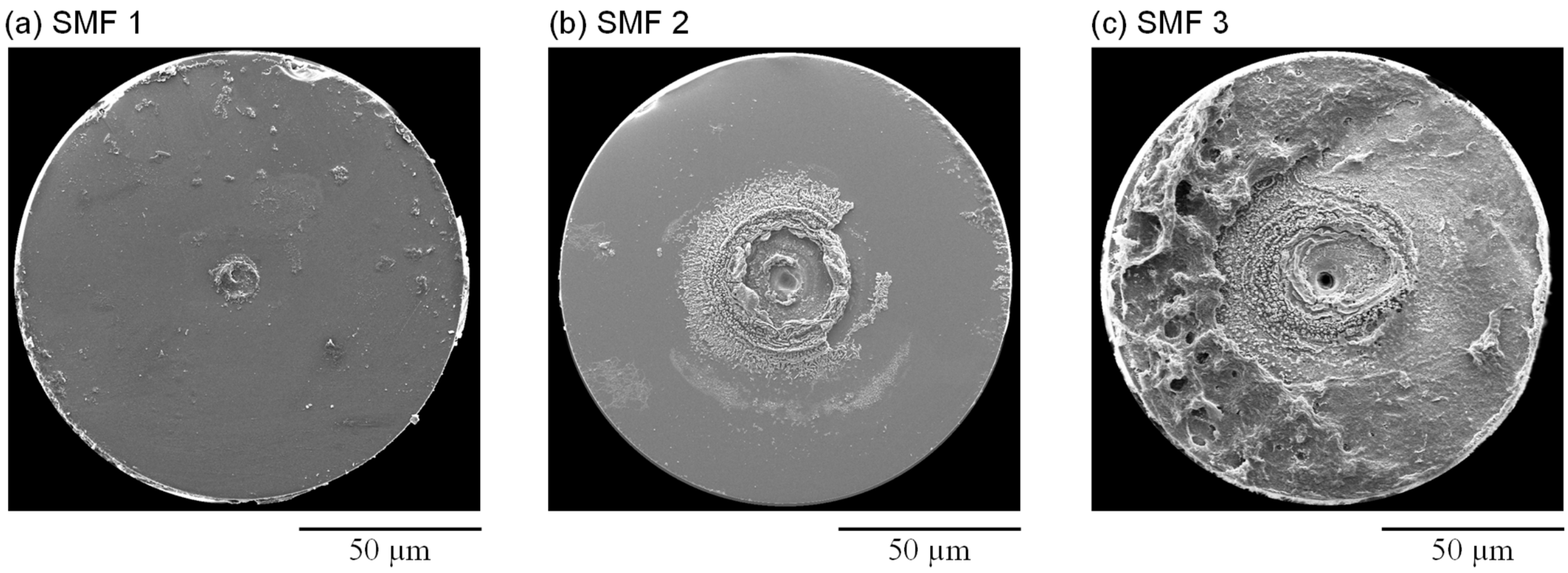

FIGURE 5. Cross-sections of the standard single-mode optical fibers, (a) SMF 1, (b) SMF 2, and (c) SMF 3, showing the vortex structures inscribed by the shock waves generated during laser exposure of 5, 10, and 20 minutes, respectively.

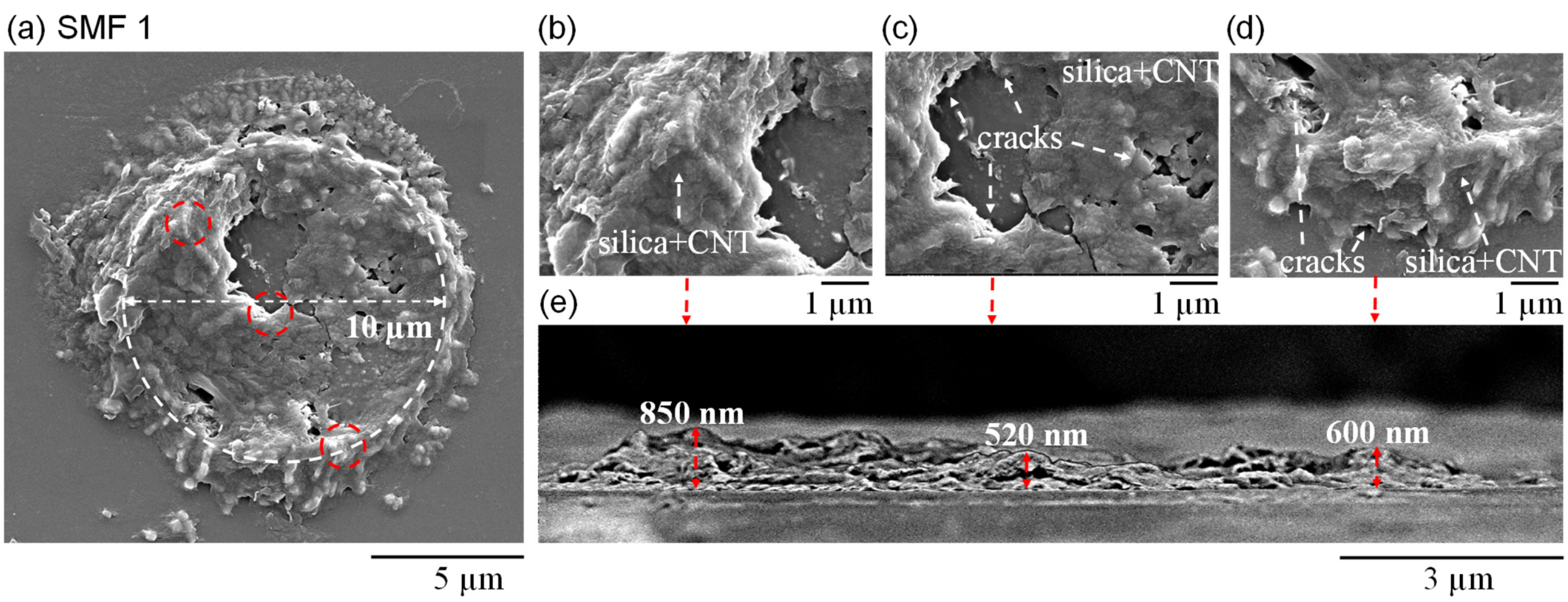

FIGURE 6. (a) SMF 1 cross-section with the CNT-silica vortex structure produced by the shock waves striking the fiber core region during 5 minutes of laser excitation. The structure exhibits a helical pattern with peaks at the borders decreasing to the central region (red circles): Details of the ring (b) upper peak, (c) central region, and (d) lower peak, with the respective (e) thickness profiles (lateral view).

The vortices increase in diameter and thickness from the fiber core center for longer laser duration. The material, geometry, and profile properties are characterized for each fiber sample. The shock wave properties ablating the Fibonacci vortices are discussed in the next section.

Fig. 6(a) shows the SMF 1 core surface with details of the carbon-silica nanostructure, composed of helical layers connected by valleys and peaks (three of them are indicated with red circles) forming an approximate 10 μm diameter ring (expanding over a 15 μm diameter). Details of the upper, central, and lower regions are shown respectively in Fig. 6(b)-(d). Fig. 6(e) shows the lateral profile around the fiber core, indicating the peaks' thickness of 850, 520, and 600 nm, respectively. The shock wave with a high-density core of nanotubes colliding with the fiber removes silica from the fiber core ($D_C$ = 8.2 μm), depositing the ablated material on the core borders and surrounding cladding (layers spread in the fiber cross-section away from the fiber core might be formed from the ablated material).

The structure inside the ring is composed of silica mixed with carbon and negligible impurities (mostly minerals, probably from the CNT solution, employed recipients, CNT powder, and fiber). Fig. 6(b)-(d) show details of internal and external cracks in the ring, suggesting that silica corrosion may be deeper than that visible on the fiber surface, potentially burying nanotubes in underground regions in the fiber core.

Fig. 7(a) shows the SMF 2 cross-section with the vortex structure. The shock wave core strikes the fiber axis, ablating a hole in the core, spreading the ablated material over the cladding. Simultaneously, the surrounding rotating vortex flow reshapes the extracted material, forming concentric rings. The first (R1) and second (R2) rings show an almost circular profile with respective diameters of about 13 and 31 μm, with peaks and valleys along the ring circumference. The fine structure around R2 spreading in the cladding indicates a third partially developed ring (R3) (the outer incomplete layers also evidence initiating rings).

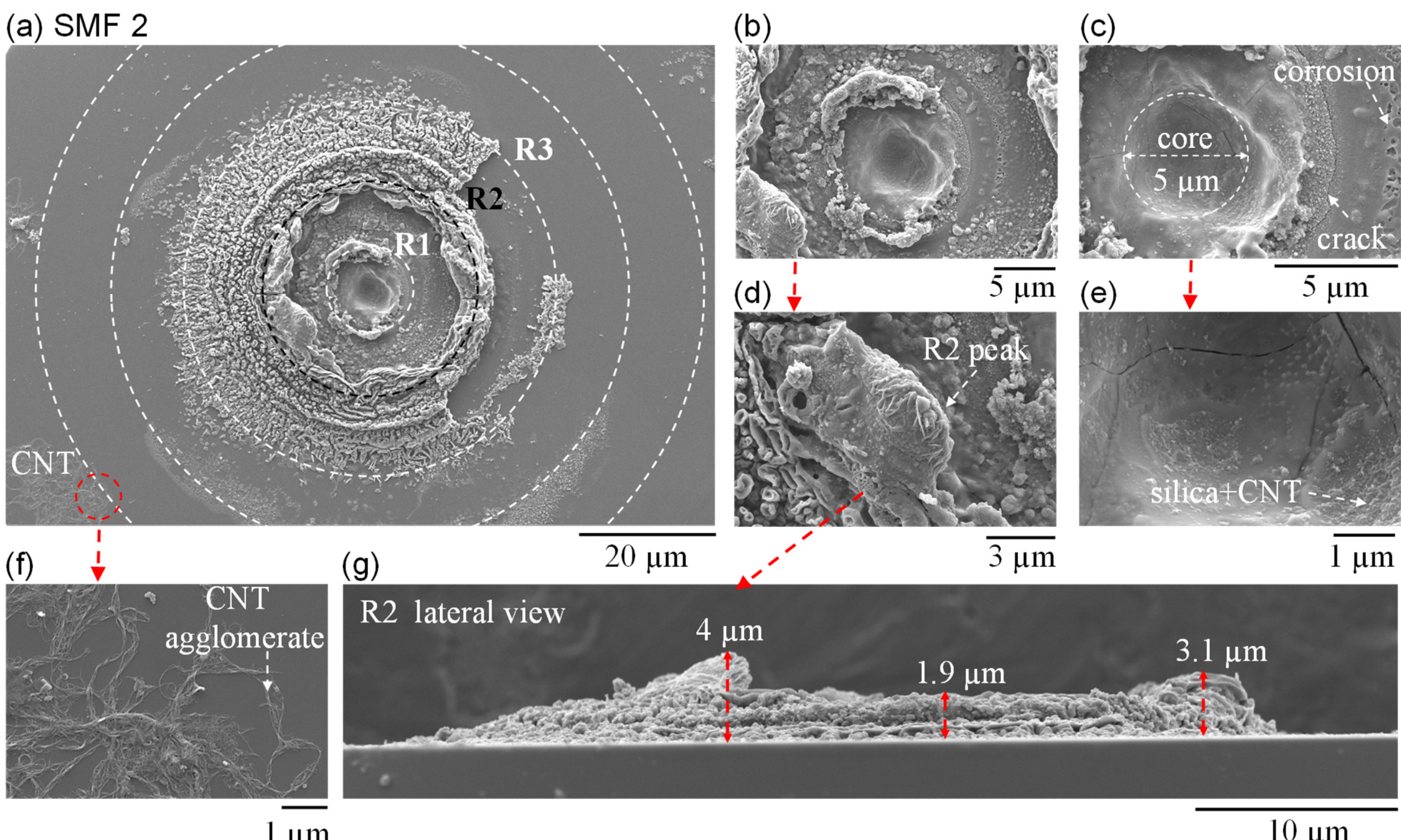

**FIGURE 7.** **(a) SMF 2 cross-section with the carbon nanotube (CNT) - silica vortex structure produced by the shock waves during 10 minutes of laser excitation. The structure shows a helical pattern forming concentric rings (R1, R2 and R3) with a central hole in the fiber core. Details of the (b) fiber core, (c) R1 valley, (d) R2 peak, (e) inside the core hole, and (f) CNT agglomerates over the cladding. (g) R2 lateral profile indicating the thickness of visible ring peaks and valley.**

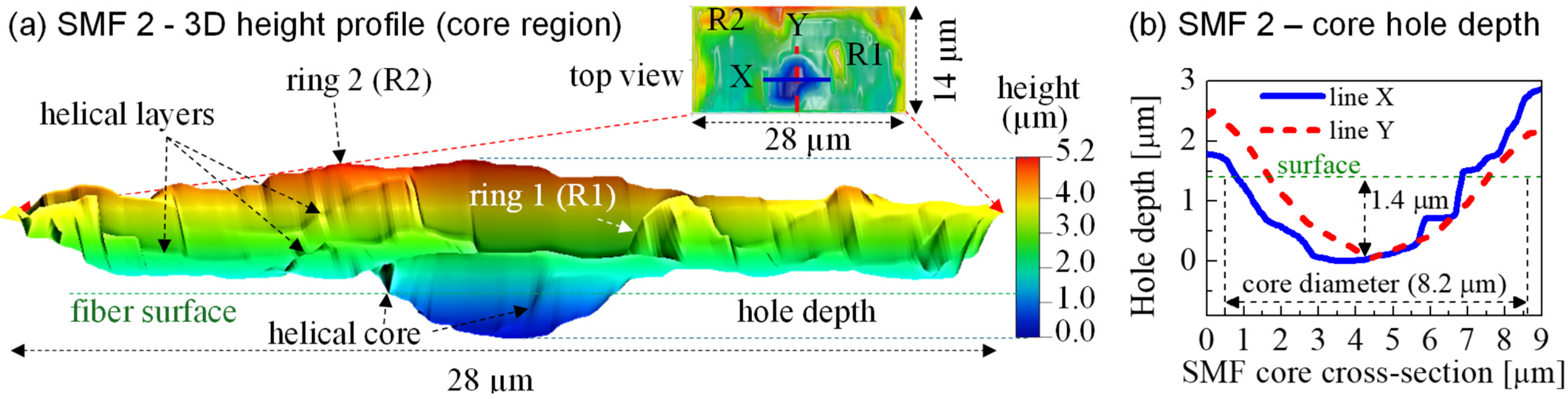

**FIGURE 8.** **(a) SMF 2 – 3D height profile of the fiber core surrounded by the first (R1) ring and partial internal region of the second (R2) ring, as indicated in the top view (inset). (b) Fiber core hole depth profile along the X and Y lines indicated in the top view.**

Fig. 7(b) and 7(c) show top view details of the fiber core and R1. The hole increases in depth, converging to the core center, with a deeper region covering about 5 µm in diameter. Fig. 7(c) shows a crack in the R1 valley, indicating the wave pressure on the ablation mechanism: the wave peaks induce high pressures, ablating the silica, while the low-pressure wave nodes confine the material, creating peaks in the ring. This is also noted for R2 with reduced ablation along the ring, caused by pressures radially decreasing from the fiber axis [18]. Fig. 7(g) shows the R2 lateral profile, indicating the thickness of the visible ring peaks (4 and 3.1 µm) and a valley (1.9 µm). The material analysis in the core region and surroundings denotes structures mostly composed of silica and carbon.

We mention that individual nanotubes deposited on the fiber surface are invisible to the practical resolution of the employed microscope (SEM). Nevertheless, nanotube agglomerates mixed with silica are observed coating the hole walls in Fig. 7(e), and over other regions in the structure and the cladding (Fig. 7(f)).

Fig. 8(a) shows the 3D height profile of the SMF 2 core surrounded by R1 and R2 (partial internal region) measured with the profilometer, as indicated in the top view (inset). This profile provides details of the core outer helical boundaries and layers deposited inside R2 (which cannot be seen considering the entire ring in Fig. 7(g)). We note that helical layers diverging from the fiber core connect the two rings. Fig. 8(b) shows the hole depth asymmetry caused by the diverging helices over the $D_C$ = 8.2 μm fiber core with a 1.4 μm maximum depth (measured along the X and Y lines in the top view).

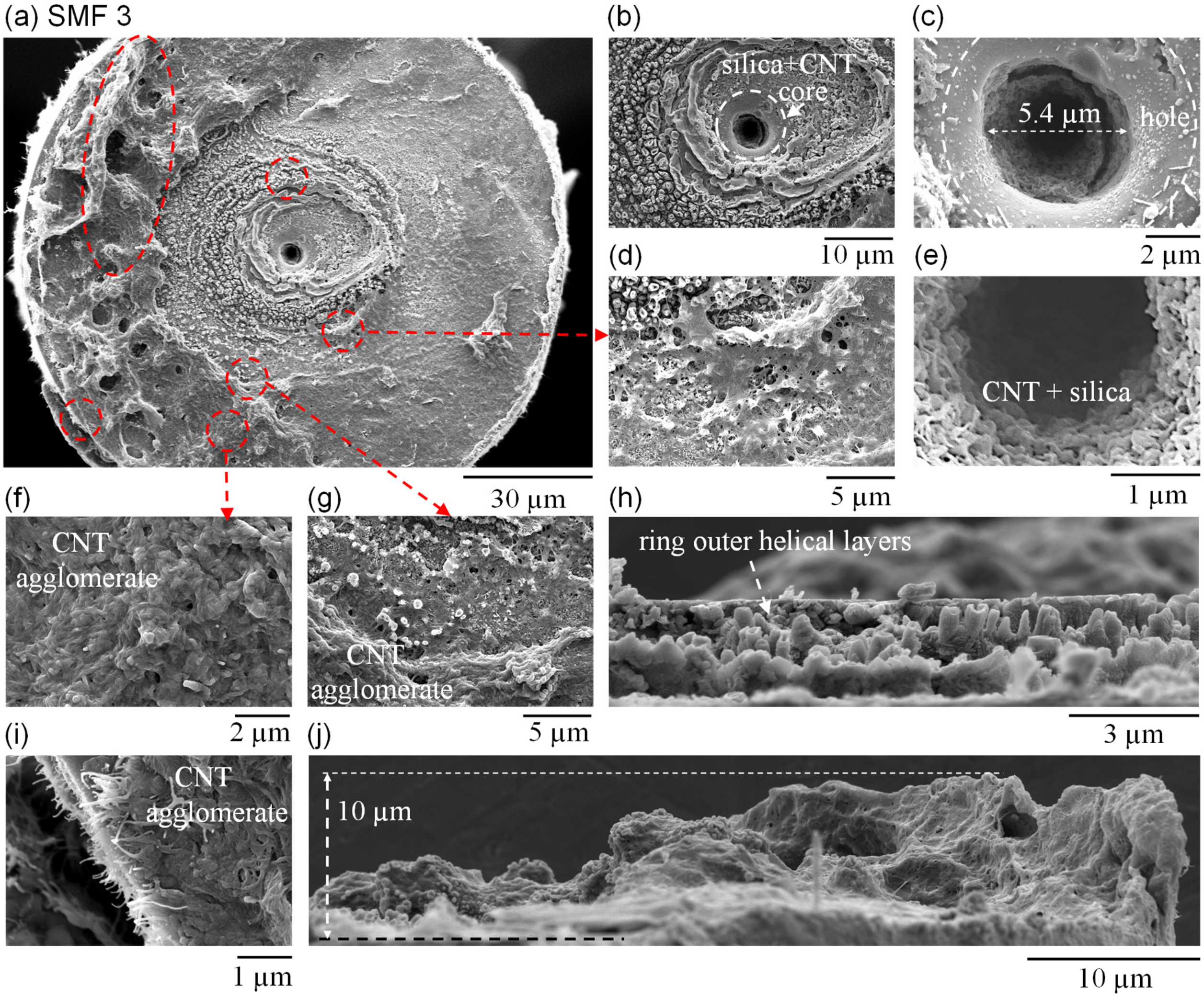

**FIGURE 9.** **(a) SMF 3 cross-section with the carbon nanotube (CNT) - silica vortex structure produced by the shock waves during 20 minutes of laser excitation, indicating details (around red curves) of the (b) fiber core and highest ring, (c) ablated hole and (e) its internal walls, (d)(g) increasing deposition of CNT bundles in the cladding, and (f)(i) dense concentration of CNT bundles agglomerating towards the fiber edge. (h) Lateral outer view of the highest ring indicating helical layers (upper red circle in (a)). (j) Lateral profile of the highest deposited CNT layer.**

Fig. 9(a) shows the SMF 3 cross-section with the vortex, indicating a printed signature of the shock wave funnel rotating and depositing particles on the fiber. The vortex waves impinge a maximum pressure along the fiber axis, ablating a valley with increasing depth to the fiber core center (hole), as shown in Fig. 9(b). Simultaneously, the radially decreasing pressure spreads and reshapes the ablated material around the core, accumulating CNT bundles on the outer fiber cladding and edges. The off-center non-circular deposited layers mainly around the core (and thickness variations on the cladding) are probably caused by a tilt of the fiber cross-section or instabilities in the vortex flow [19], [30].

Overall, Fig. 9(c) and 9(e) show an apparently circular central hole with a deeper region over a 5.4 μm diameter, indicating that the fiber core is less affected by asymmetries than the cladding and edges.

The deposited layers are highly concentrated in carbon mixed with silica, coating the hole walls in Fig. 9(e). The valley and rings near the fiber core are composed of silica mixed with carbon and negligible minerals, contributing to forming the granular structure shown in Fig. 9(b) (left side). Fig. 9(h) shows the lateral view of the outer granular layers around the ring indicated by the upper central circle in Fig. 9(a). These layers gradually overlap with nanotube bundles spreading on the fiber cladding surface (Fig. 9(d) and 9(g)) and densely agglomerating towards the fiber edges (Fig. 9(f)).

Fig. 9(i) shows CNT bundles radially aligned on the fiber edges and lateral surface, indicating the effect of the expanding vortex flow around the fiber tip. Moreover, CNTs deposited in the denser and thicker layers over the fiber cross-section exhibit helices with peaks and valleys, as observed in the highest layer in Fig. 9(j) (the lateral profile indicates the highest peak with about 10 μm thickness). Peaks and valleys of multiple deposited helices overlap on the fiber surface, forming cavities, mostly observed on the left side in Fig. 9(a).

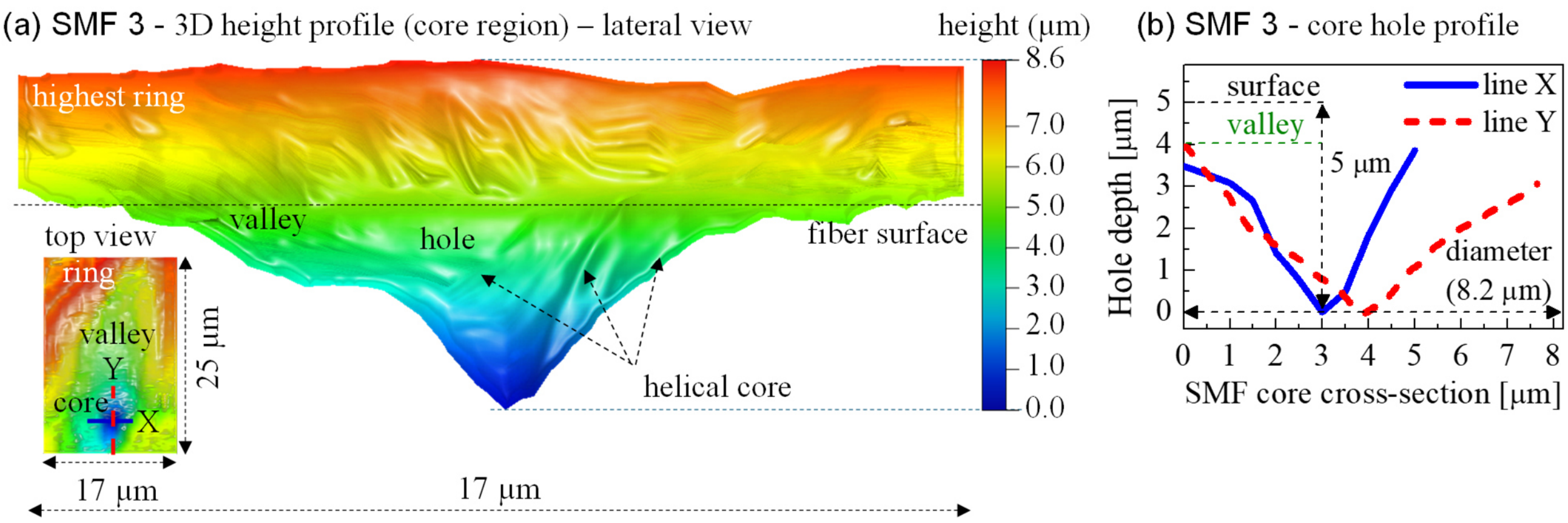

**FIGURE 10.** **(a) SMF 3 – 3D measured height profile of the fiber core surrounded by the highest ring border, as indicated in the top view (inset), showing the ablated silica region forming a valley with increasing depth in the hole. (b) Hole depth profile along the X and Y directions in the fiber core diameter (8.2 μm) as indicated in the top view in (a), showing helices converging to the core center with a 5 μm maximum depth.**

Fig. 10(a) shows the 3D height profile of the fiber core surrounded by the highest ring border measured with the profilometer (as indicated in the top view inset). This lateral profile reveals internal details of the ring, ablated valley, and the hole in the fiber core. We note helices around the hole converging at its center, inducing an asymmetric profile along the X and Y directions, as shown in Fig. 10(b) (indicated in the top view in Fig. 10(a)). The hole covers the fiber core diameter (8.2 μm) with a 5 μm maximum depth.

### *B. ANALYSIS AND PROPERTIES OF THE VORTEX SHOCK WAVES*

Fig. 11 shows optical emissions in the visible spectral range from high-density clouds of carbon nanotubes under laser radiation, indicating the axial development of vortex shock waves generated with SMF 3 (fiber is not visible and has been drawn to scale for visual guidance only). The visualization of wave emissions takes a few minutes after the laser is ON and gradually increases over time. The shock waves are monitored during short periods, showing distinguished cylindrical vortex, toroidal, and cavitation flows (as previously illustrated in Fig. 1(g)).

Fig. 11(a)-(f) show selected images of a developing vortex monitored for 2 minutes, with rotating helices forming two visible rings (other unseen rings might exist along the container). The rings decrease in diameter from 1 mm to a minimum at the fiber surface (the bright circular region on the fiber might indicate the wave starting edge, as previously seen in Fig. 1(b)). Fig. 11(b) shows the evolving helices along 1.3 mm, completing almost a full rotation with a 540 μm spatial wave period (Fig. 11(c)). Fig. 11(d) shows particles flowing from the larger to the smaller ring, indicating CNTs depositing over the fiber tip. This is followed by increasing toroidal waves propagating perpendicularly to the vortex, as shown in Fig. 11(e). Afterwards, the accumulating nanotubes in the vortex merge the rings, engaging particles in the whole vortex flow (Fig. 11(f)). Consequently, CNTs deposited on the fiber radially expand the acoustic source, inducing strong brightness around the fiber tip.

Fig. 11(g) shows strong CNT cavitation forming a bubble along the fiber axis, monitored for about 1 minute (as previously observed for shock waves with dye absorbers [17], [18], [19], [23]). Fig. 11(h) and 11(i) show the CNT particles colliding as a jet with the fiber, forming a high-density core converging as a funnel towards the fiber core (as shown in the inset in Fig. 11(i)). The CNTs show apparently stationary flows approaching the container walls. Afterwards, the brightness from the CNTs was too high, and the wave flows could no longer be characterized, indicating that the wave dimensions might become larger, expanding and confining over the whole syringe container.

The shock wave properties are estimated employing the formulation and parameters previously described in Section II.B. The vortex jet strikes the CNT bundles deposited on the fiber core surface with axial velocities approaching $V_P = 725$ m/s, igniting internal shock waves in the nanotube bundles with hypersonic velocities of $U_S = 5742$ m/s (the estimated values agree well with results summarized in [42]). We mention that the estimated CNT velocities in the fluid are significantly higher than those achieved by thermally induced fluid convection, thermophoresis, and thermocapillary effects ($V_P = 1$ μm/s maximum reported) [14], indicating that acoustic and shock wave mechanisms strongly dominate over other thermally induced effects in the CNT solution [60].

The CNT shock waves impinge pressures on the silica of about $P_S = 6.7$ GPa, partially reflecting and inducing tensile stresses with magnitudes higher than the silica strength, usually in a range up to $\sigma = 5.72$ GPa [61], [62]. For SMF 1 and 2, the shock wave pressures are high enough to eject particles and prevent CNTs from bundling in the core, resulting in apparently dense silica-carbon layers composing the rings. In contrast, for longer ablating periods (SMF3 - 20 minutes), the particles accumulating in the hole are unable to escape from the fiber core, agglomerating and coating the core walls.

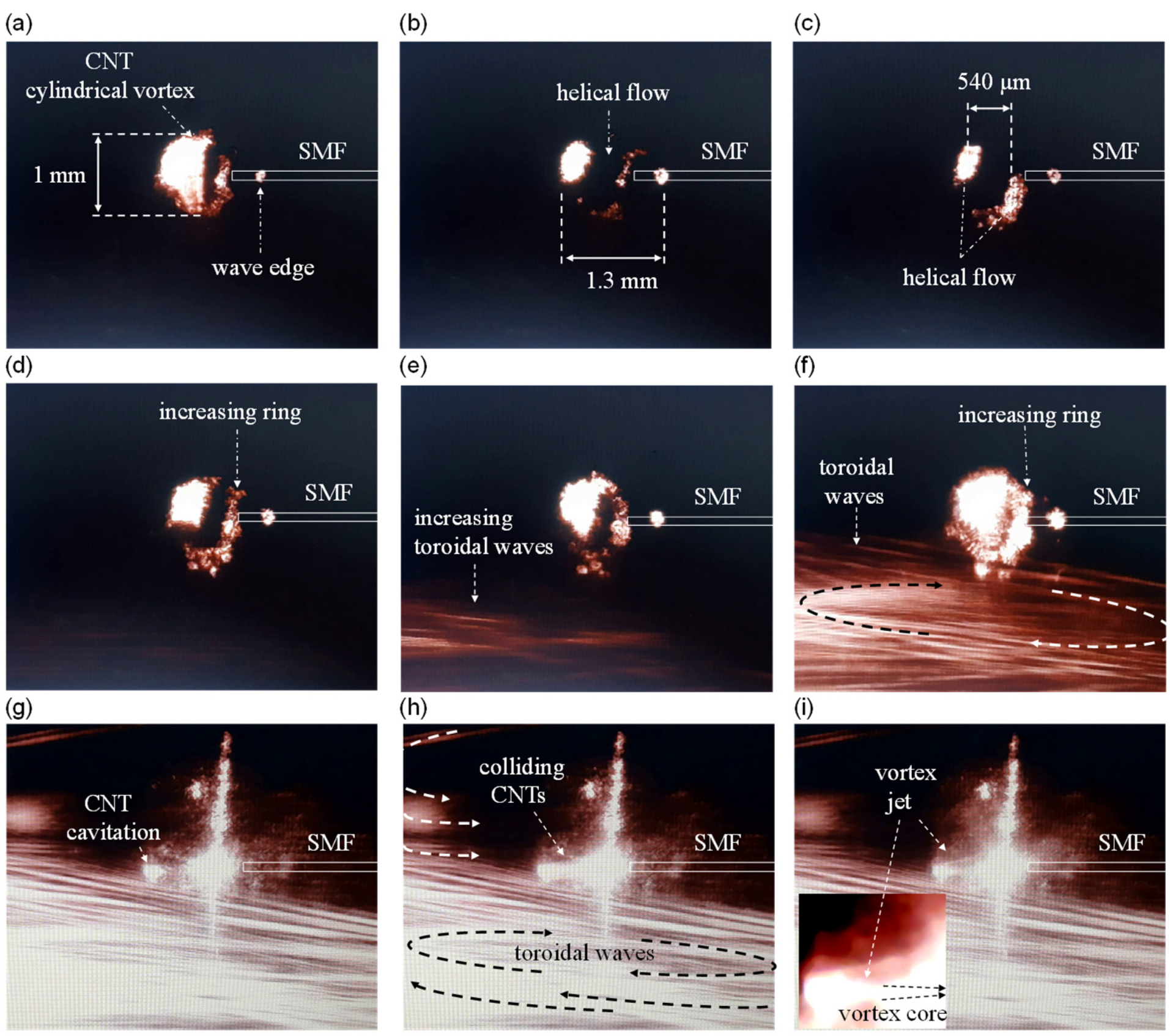

**FIGURE 11. Optical emissions in the visible spectral range showing the lateral view of developing vortex shock waves composed of clouds of carbon nanotubes (CNT) dispersed in methanol under laser radiation from a standard single-mode optical fiber (SMF 3 sample): (a)-(f) vortex flow developing (g) CNT cavitation further (h) striking particles as a (i) jet to the fiber, showing details of the vortex core (inset).**

We state that the used infrared laser power at 980 nm is undetected for the employed microscope and thus invisible in the images in Fig. 11. Nevertheless, laser power spreading in the solution and reflecting at the solution-container interfaces is expected. In addition, CNT bundles absorbing and radiating partially the laser energy might energize adjacent bundles in the flow, contributing to enhancing wave expansion even at distances far away from the fiber.

Overall, the observed vortex, toroidal, and cavitation flows should be present with distinct magnitudes in all stages shown in Fig. 11, with increased visibility in regions where particles are mostly dense and energized. These wave patterns are also expected for SMF 1 and SMF 2, where the vortex flows could not be clearly identified due to shorter laser durations and reduced visibility, suggesting that nanotubes accumulating for longer laser durations contribute to improving visibility of the developing vortices.

The increased interaction of highly concentrated CNTs, high laser power, and fiber in the syringe cavity is essential to generating shock waves, engaging more particles in the acoustic flow, and particle cavitation (bubbles). Nevertheless, for reducing CNT concentrations and powers, the required pressures for silica ablation may not be achieved, and the employed setup will be useful for depositing nearly planar CNT layers on fiber tips [56]. In addition, changes in the CNT mechanical properties should influence the shock wave speeds and pressures (Eq. (6)). The estimated CNT bundle velocity, $c_{CNT}$ = 4800 m/s, is based on a bundle Young's modulus of $E_{CNT}$ = 37 GPa, which approaches experimental values considering defects and non-uniformities commonly observed in the nanotube arrays in practical devices [31], [46]. We note that $E_{CNT}$ is significantly lower than that of individual nanotubes, around 230 GPa [46], suggesting that increasing $E_{CNT}$ should enhance shock velocities and pressures.

Previous studies have shown that vortices print their boundaries with particles accumulating on solid surfaces, providing detailed information about the acoustic shock wave properties [63], [64]. Fig. 12 shows SMF 1, 2, and 3 with the CNT-silica vortices, indicating details of evolving helices from the fiber core. Fig. 12(a) shows the SMF 1 printed vortex indicating the superposition of three helices gradually increasing in thickness from the ring valleys (blue circle) to the peaks (red circle). The helices' thickness variation is observed on both SMF samples and is caused by the helical phase distribution in the vortex front colliding with the fiber [54], [65]. The helices rotate, inducing a circular flow, ablating and depositing the material while forming the ring ($R_V = 5$ µm) covering the fiber core ($R_C = 4.1$ µm) [52].

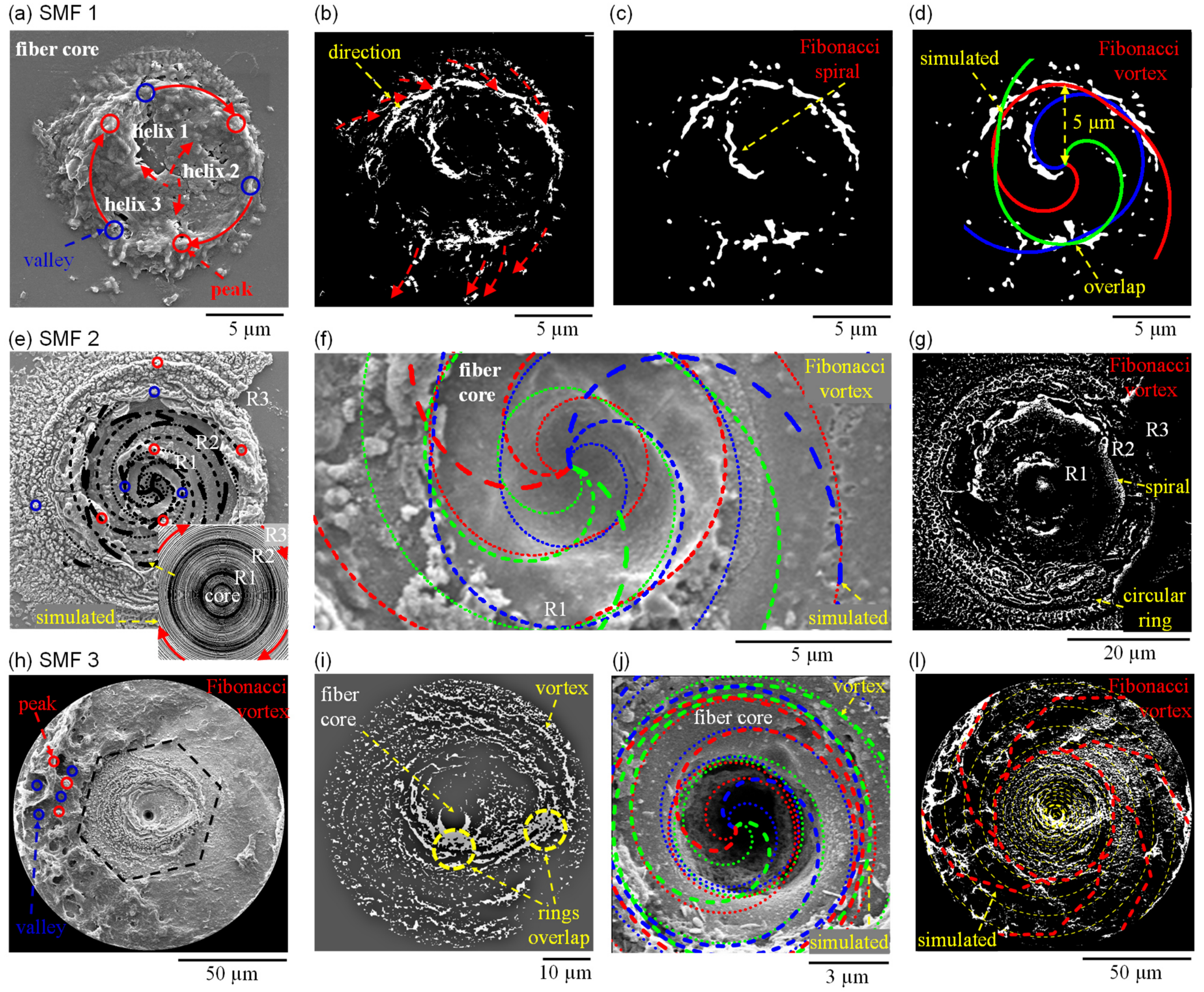

**FIGURE 12.** Shock wave inscribed Fibonacci vortices on standard single-mode optical fibers (SMF): (a) SMF 1 indicating peaks (red circle) and valleys (blue circle) caused by three merged vortex helices. SMF 1 binary images showing the (b) vortex rotation direction and (c) a Fibonacci spiral printed in the fiber core. (d) The measured vortex is compared to the analytical simulation. (e) SMF 2 vortex compared to the simulation (inset shows the rings over-rotating, indicating the measured diameters). (f) Details of the SMF 2 core compared to the simulation showing the development of three Fibonacci helices. (g) Binary image of the measured vortex reveals the helices merging into a unique spiral and depositing particles in a circular pattern. (h) SMF 3 Fibonacci vortex forming a hexagon with overlapping peaks and valleys, inducing dense CNT cavities in the cladding. (i) FFT image of the vortex around the fiber core showing layers overlapped by asymmetries. (j) The Fibonacci vortex ablates the hole while depositing mixed carbon and silica layers following Fibonacci spirals. (l) Binary image compared the simulations, indicating decreasing vortex rotation from the fiber core to the cladding and edges.

Fig. 12(a) is image-processed by employing a combination of binary (black - white) functions (e.g., threshold, erode, median) and fast Fourier transform (FFT), which are available with the ImageJ (FIJI) software [59]. Fig. 12(b) and 12(c) show the vortex, respectively with binary and erode-median functions, enhancing details of the vortex rotation direction. Fig. 12(c) reveals a Fibonacci spiral, which is compared to the analytical simulation in Fig. 12(d) (*xy* axes are omitted for improved visualization).

We note in Fig. 12(d) that the simulated spirals overlap in the regions of large material deposition on the vortex ring, indicating high agreement of measured and simulated vortices.

SMF 2 and 3 are characterized by using similar methods and simulation parameters of SMF 1. Fig. 12(e) shows the SMF 2 vortex with valleys and peaks in the rings (R1, R2, and R3). The simulation (black curves) projects the rotating three helices inside ($D_{V1} = 5$ µm) and outside ($D_{V2} = 10$ µm) the fiber core, approaching R2 ($D_{V3} = 24$ µm) (the inset illustrates the helices over rotating forming concentric ideal circular patterns). Fig. 12(f) shows the SMF 2 core vortex compared to the simulated Fibonacci spirals. We note high agreement of the measured vortex boundaries inside the fiber core with the simulated spirals. Fig. 12(g) highlights the measured CNT-silica layers deposited between the rings, confirming that the three helices merge as a unique helical flow, which rotates, depositing particles circularly over the fiber cladding.

The Fibonacci vortex is highly developed in SMF 3, as shown in Fig. 12(h) (valleys and peaks intersecting out of phase induce cavities in the dense CNT layers - left side). In addition, radially decreasing vortex rotation from the core forms a nearly hexagon-shaped distribution on the cladding [52]. The helices diverge from the core center and merge into helical rings expanding around the core, as shown in Fig. 12(i) (FFT) (some rings overlap due to vortex asymmetries). Overall, agreement of measured and ideal simulated vortices in Fig. 12(j) shows that the vortex asymmetries are negligible in the fiber core. We note that the shock waves remarkably ablated the hole, while depositing carbon nanotube layers (300 nm average thickness) in the fiber core walls following the Fibonacci spirals. Fig. 12(l) shows the binary SMF 3 vortex compared to the simulation with radially decreasing rotation [53], indicating that the measured shock waves form compound vortices with a forced core rotating with high azimuthal velocities, surrounded by a free flow usually named as irrotational [26], [33], [34], [53].

We conclude that the generated Fibonacci vortex shock waves, inscribing CNT-silica vortices in both SMF samples, follow the same physical mechanisms, with increasing radial and thickness dimensions over time. Nevertheless, significant changes in laser power, CNT-solution, syringe cavity dimensions, fiber tilts, and misalignments should cause vortex instabilities and thickness variations in the ablated structures, mainly in the cladding, and for long laser periods, preventing the deposited layers from achieving a helical or circular profile. Overall, the results show that these variations should be decreased in the fiber core, due to the higher velocities and pressures imposed by the shock waves.

### *C. FINAL CONSIDERATIONS*

Some details and challenges should be considered for proper fabrication and characterization of the shock waves and vortex devices, mainly regarding the CNT type, repeatability and sensitivity.

We used SCNTs due to their unique capability to adjust the spectral absorption band peaks by choosing the nanotube diameter. In particular, the employed SCNTs exhibit high optical absorbance peaks at wavelengths matching the employed laser at 980 nm, thereby improving efficiency for laser absorption, thermal expansion, and generation of shock waves. In addition, the used SCNTs offer broad absorption around 1100 nm, which is promising for applications in Ytterbium-doped fiber lasers and photoacoustic ultrasonic devices usually operating around 1060 nm. In general, SCNTs have shown lower phase noises, higher modulation depths, and higher saturation intensities compared to graphene in pulsed fiber lasers, resulting in increased mode-locking thresholds and pulse energies, as well as wider laser optical spectra [10].

The SMF tips (with the vortices) were cut into small pieces about 4-5 millimeters in length to fit vertically in the SEM's standard metallic stubs. The SMFs are fixed to the stubs by using strong metallic and carbon tapers. This procedure was required to image high-quality nanoscale details of the vortices, allowing the stubs to be further adapted to the profilometer and Raman spectrometer, facilitating the characterization of dimensions and material of the CNT-silica layers. Nevertheless, the metallic stubs and conductive tapers were not compatible with other general laboratory equipment, making it not suitable for thermal and mechanical tests. Additionally, the individual short fiber pieces are elaborate to handle and fix, making additional mechanical and thermal characterization too difficult. Further studies might measure mechanical and thermal stability of the vortices before SEM measurement or by using stubs adapted for long fiber lengths, allowing the samples to be applied in devices and further characterized in other equipment. Overall, we mention that our SMF vortices were characterized by a contact profilometer using a diamond probe directing pressuring the silica-carbon layers on the fibers during the height scan. Although it is not shown, the deposited layers remained conserved after profile characterization, indicating good mechanical stability of the CNT-silica layers.

The shock velocities and pressures are estimated at the interface of the deposited CNT layers and the silica cross-section. These shock waves are generated inside the nano CNT layers in the core region, as shown in the SEM images. Confirming the existing literature, the shock waves might propagate in the nano layers with sub nano spatial wavelengths and gigahertz frequencies. The main experimental challenges associated with direct velocity and pressure measurements are listed below:

a) The fiber tip is confined inside the syringe making its access difficult to other external devices and equipment. Alternatively, another microscale sensor device (as a sensing fiber) could be inserted in the syringe with the SMF sample to monitor the fiber tip but could affect or be affected by the fluid vortex generation. Furthermore, this additional sensor might still not access the CNT-silica interface as it ablates dynamically getting deeper over time.

b) The shock wave parameters might be directly measured by monitoring the reflected power at the SMF tip, considering it as a nano Fabry-Perot cavity reflecting at the silica interface and at the CNT external interface. For an ideally flat non-ablated SMF-CNT interface, the ultrafast shock waves might require compatible high-speed amplified photodetectors and oscilloscopes, and perhaps a specific coupled wavelength and signal processing method to resolve the reflected signals. Nevertheless, the shock waves dynamically change the silica-CNT cavity from an initially flat interface to a vortex hole patterned surface, increasing absorption and spreading optical power in distinct directions, becoming critical reflection monitoring. Further studies overcoming these challenges might provide useful real-time monitoring of shock waves and vortex structures.

c) The shock velocities might be indirectly estimated by externally monitoring the SMF tips using high-speed cameras

through a transparent glass syringe. Thus, the lower particle velocity ($V_P$ = 725 m/s) in Eq. (6) might be measured by monitoring the CNT bundles moving positions in time towards the fiber tip [17]-[19]. This technique requires high image resolution and advanced image-signal processing to properly identify particles trajectories. We hope that our proposed method, using particle visible emissions, might be useful in this direction in combination with analysis software, such as ImageJ (FIJI - Trackmate).

The vortices analysis and their comparison with ideal simulations indicate high experimental repeatability, showing progressive ablation in the SMFs over time. The setup should be calibrated before the experiments by aligning the XYZ positioners, the syringe and fiber holders (V-grooves), and the SMF tip to the syringe aperture using test fibers. In this stage, the microscope focus on the SMF tip is also adjusted. During the experiment, the aligned components and loaded syringe (cavity geometry and CNT solution) are fixed, and fiber samples can be replaced in the holders. Additionally, repeatability relies on the fixed laser power and fabrication quality of the fiber cross section, its preparation (cleavage and cleaning) and replacement in the holder preventing tilts and bends.

Regarding parameter sensitivity, the use of multiple samples might deploy nanoparticles in the CNT solution. Favorably, we have not observed decreased performance for the three employed samples, probably because of the initially high CNT concentration and sufficient solution volume (0.2% maximum change of CNT concentration should cause negligible variations in solution viscosity and density, and the resulting particle velocity [21],[35]). In addition, cavity variations satisfying, $H/R \geq 4$, should also not affect shock wave flows (as well as increasing laser power up to 3%). However, significant decrease in carbon concentration, cavity geometry (reducing solution volume) and laser power, might reduce shock wave performance, ablation capacity and nanotube deposition. Besides, fiber cleavage angles, tilts and misalignments might induce irregular laser distribution in the solution, and radially asymmetrical CNT deposition and ablation on the fiber, mainly in the cladding surface. Overall, syringe confinement, high laser power, high CNT concentration in methanol, and deposition time, are the dominant influencing factors to achieve ablation and robust vortex structures. Further studies might investigate the lower and upper thresholds values for the considered parameters, indicating quantitative ranges for CNT deposition and ablation.

## V. CONTRIBUTIONS AND OUTLOOK

This paper reviews extensive and important literature in distinct scientific areas, sharing similar approaches and physical mechanisms, to provide the most accurate information about the generation and application of the investigated vortex shock waves. Thus, to the best of our knowledge, our study provides the following significant contributions:

(a) First demonstration of fabricated CNT-silica Fibonacci vortices with controlled radial and thickness dimensions on optical fiber tips. The inscribed Fibonacci vortices contain holes decreasing in diameter to the core center while increasing in depth (5 µm maximum), depositing nanolayers of carbon nanotubes and silica following Fibonacci spirals (no reference has been found with similar features).

(b) First experimental demonstration of laser-induced vortex shock waves formed by single-walled carbon nanotubes with standard single-mode optical fibers. The shock waves show estimated hypersonic velocities (5742 m/s) and high pressures (6.7 GPa) overcoming the silica's tensile strength (Fig. 11 shows vortex flow dimensions from 5 to 15 times larger than those reported using dye absorbers and large multimode fibers [17], [18], [19]).

(c) The estimated CNT velocities in the fluid (725 m/s) are significantly higher than those observed with fluid convection, thermophoresis, and thermocapillary mechanisms ($V_P$ = 1 µm/s) [14]. In addition, the CNT particles focusing on the fiber core in Fig. 11(i) overcome critical laser radiation pressures that deviate nanoparticles from the fiber, preventing CNT deposition in the core. Thus, the inscription of the vortex highly concentrated in the SMF 1 core confirms that particles can be focused and deposited on the core region without using polymers in the solution, reducing cost while increasing power damage threshold [1], [2], [3], [4], [15].

(d) The CNT-silica layers deposited in the SMFs core show thicknesses from less than 300 nm to 850 nm, which are the thinnest compared to previous devices using laser-induced methods, dip coating, and wallpaper techniques (900 nm – 49 µm) [1], [2], [3], [4], [10], [12], [15].

(e) The syringe-based cylindrical cavity strongly confines the acoustic fields, contributing to generating the shock waves, reducing consumed energy, while providing a simple and cheap alternative to CNT deposition. These benefits are significantly increased when considering silica ablation employing high-power femtosecond lasers [66], [67]. In addition, the 1 mL solution volume used to fabricate three vortex samples is much lower compared to techniques usually employing Beckers [9], consequently reducing material, device size, weight, and overall cost. These benefits should have a higher impact when using expensive optical absorbers such as silver and gold.

For future development of this research, we hope that the demonstrated vortices, including ring-shaped CNT distribution around the fiber core, should enhance absorption of evanescent optical fields, reducing thermal damage and non-saturable losses in the fiber core, while increasing power output and efficiencies of ultrafast mode-locked fiber lasers [1], [6], [10]. For fiber-based optical sensors, the CNT layer deposited on the fiber tip should benefit from the carbon's ability to absorb molecules of environmental measurands, modulating the nanotube's characteristics (such as electrical, geometric, and optical properties), being suitable for chemical, biological, and gas sensing [68]. Furthermore, the fabricated Fibonacci vortices should be investigated to modulate the phase of the propagating optical modes coupled to the fiber, enabling fiber-based spatial light modulators (SLM) for the generation of angular orbital momentum (AOM) in demanding photonic applications, including optical communications, quantum optics, and optical micromanipulation [54], [65], [69].

In summary, the demonstrated devices are promising to increase confinement and spatial resolution of biomedical photoacoustic devices, improving propagation and focus of ultrasonic pulses to reach microscale biological targets [70]. Further studies might consider the fabrication of vortex structures employing distinct CNT types, fluids, laser parameters, and optical fibers, such as multimode fibers, photonic crystal fibers, microstructured optical fibers, and multicore fibers [68]. Thus, CNTs might be deposited inside the fibers with large air holes, such as suspended core fibers and antiresonant hollow core fibers [71], [72], potentially leading to the development of innovative in-fiber vortex devices.

## VI. CONCLUSION

We have experimentally demonstrated laser-driven shock waves based on single-walled carbon nanotubes that ablate vortices in standard single-mode optical fibers. Three SMF samples are investigated by injecting a high-power laser into a CNT-methanol solution for periods of 5, 10, and 20 minutes.

The syringe-based cavity confines the thermally induced acoustic fields in the fluid, generating shock waves with velocities (5742 m/s) and pressures (6.7 GPa) overcoming the fiber tensile strength, ablating fiber tip surface and depositing CNT-silica layers with vortices pattern. The shock waves and inscribed structures are characterized, revealing vortices, respectively, for SMF 1, 2, and 3, increasing in diameter (approximate maxima - 15, 54, 125 µm) and deposited thickness (850 nm, 4 µm, 10 µm) with increasing laser duration from 5 to 20 minutes. Measured and simulated results confirm that the shock waves and ablated vortices are formed by rotating Fibonacci helices, drilling holes in the fiber's core with a 5 µm maximum depth, while deposing carbon-silica nanolayers on the core walls.

These achievements point out a new route for controlled fabrication of nano-to-micro vortex devices on fiber tips, promising the development of photonic spatial phase modulators for generating optical vortex beams and optical orbital angular momentum (OAM). These features can potentially enable novel fiber sensors, modulators for pulsed fiber lasers, and ultrasonic shock wave transmitters for biomedical applications.

## ACKNOWLEDGMENT

We thank Prof. Cristiano M. B. Cordeiro with the Laboratório de Fibras Especiais & Sensores Ópticos (LaFE) for the support with the experimental setup, and Prof. João P. V. Damasceno with the Group of Nano Solids (GNS) for providing the CNT solution. We thank the groups with LAMULT (Rosane Palissari, Bruno Camarero), LIMicro – IQ (Hugo C. Loureiro and Ana Letícia M. da Fonseca), and ITA - LPP (Dr. Nilton F. A. Neto, Dr. Isabela M. Horta, and Prof. André L. de Jesus Pereira) for the support with the characterization of the SMF samples, including, respectively, Raman spectroscopy, SEM images, and 3D profilometry.

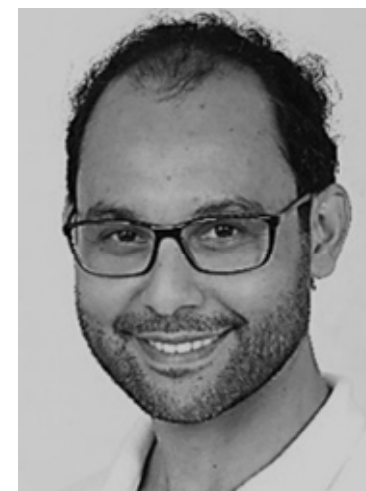

**RICARDO E. DA SILVA** received the B.S. degree in telecommunication engineering from the Assis Gurgacz University Center, Cascavel, in 2009, M.S. and Ph.D. degrees in electrical engineering from the Federal University of Technology, Curitiba, in 2011 and 2015, and the Ph.D. degree in technical physics engineering from Friedrich Schiller Universität, Jena, Germany, in 2015. From 2013 to 2015, he was a Researcher with the Leibniz Institute of Photonic Technology (IPHT), in Jena, and a Marie Curie Fellow with the Aston Institute of Photonic Technologies (AIPT) from 2019 to 2021. He is currently a Postdoctoral Researcher with the University of Campinas (UNICAMP) and a Visiting Fellow with Aston University (AIPT) in UK. His research interests include optical fibers, acousto-optic devices, photoacoustic fiber sensors, and fiber lasers for applications in telecommunications, industry, and biomedicine.

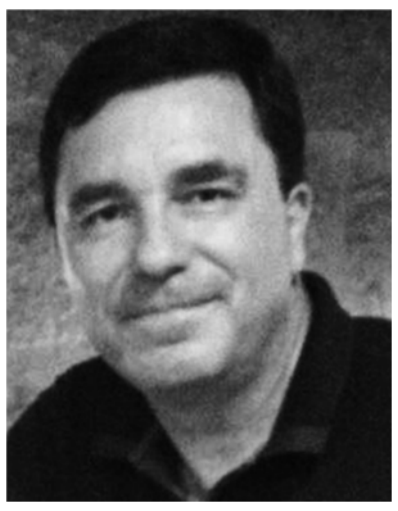

**MARCOS A. R. FRANCO** received the B.S. degree in physics from the Pontifical Catholic University of São Paulo (1983), M.S. in physics from the University of São Paulo (1991), and Ph.D. in electrical engineering from the University of São Paulo (1999). Researcher of the Institute for Advanced Studies (IEAv) and Professor at the Technological Institute of Aeronautics (ITA). From 2018-2020 was an Associated Editor of Journal of Microwaves, Optoelectronics and Electromagnetic Applications (JMOe). The main research areas are: Applied electromagnetics, Computational modeling of Electromagnetic devices, optical fibers, fiber optic sensors, integrated optics, photonic crystal fibers, microstructured optical fibers, wave propagation, microwave, and terahertz waveguides.